%% file: EXO-11-024_temp.tex
\pdfoutput=1

\documentclass[11pt,twoside,a4paper,cmspaper,final,collab]{cms-tdr}
\def\svnVersion{166724}\def\cmsCernNo{2012-103}\def\cmsCernDate{\today}\def\cmsMessage{Submitted to the Journal of High Energy Physics}

\begin{document}\cmsNoteHeader{EXO-11-024}

\hyphenation{had-ron-i-za-tion}
\hyphenation{cal-or-i-me-ter}
\hyphenation{de-vices}
\RCS$Revision: 122895 $
\RCS$HeadURL: svn+ssh://svn.cern.ch/reps/tdr2/papers/EXO-11-024/trunk/EXO-11-024.tex $
\RCS$Id: EXO-11-024.tex 122895 2012-05-20 16:53:49Z cleonido $

\providecommand\Wprime{\PWpr\xspace}
\providecommand\WPL{\ensuremath{\cmsSymbolFace{W}^\prime_\mathrm{L}}\xspace}
\providecommand\WPR{\ensuremath{\cmsSymbolFace{W}^\prime_\mathrm{R}}\xspace}
\providecommand\Zprime{\cPZpr\xspace}
\providecommand\sm{standard model\xspace}
\providecommand\MT{\ensuremath{M_\mathrm{T}}\xspace}
\providecommand\MTlower{\ensuremath{M_\mathrm{T}^\mathrm{min}}\xspace}
\providecommand\ET{\ensuremath{E_\mathrm{T}}\xspace}
\providecommand\pT{\ensuremath{p_\mathrm{T}}\xspace}
\providecommand\invpb{\pbinv}
\providecommand\invfb{\fbinv}
\providecommand\WprimeMuNu{$\Wprime \rightarrow \mu\nu $\xspace}
\providecommand\WprimeENu{$\Wprime \rightarrow \Pe\nu $\xspace}
\providecommand\WprimeKK{\ensuremath{\cmsSymbolFace{W}_\mathrm{KK}}\xspace}
\providecommand\WprimeKKn{\ensuremath{\cmsSymbolFace{W}^\mathrm{n}_\mathrm{KK}}\xspace}
\providecommand\WprimeKKtwo{\ensuremath{\cmsSymbolFace{W}^\mathrm{2}_\mathrm{KK}}\xspace}
\providecommand\WMuNu{$\PW \rightarrow \mu \nu $\xspace}

\cmsNoteHeader{EXO-11-024} % This is over-written in the CMS environment: useful as preprint no. for export versions
\title{\texorpdfstring{Search for leptonic decays of \Wprime bosons in pp collisions at
  $\sqrt{s}=7\TeV$}{Search for leptonic decays of W' bosons in pp collisions at sqrt(s)=7 TeV}}

\date{\today}

\abstract{
A search for a new heavy gauge boson \Wprime decaying to an electron or muon, plus a low mass neutrino, is presented.
This study uses data corresponding to an integrated luminosity of 5.0\invfb, collected using the CMS detector in pp collisions at a centre-of-mass energy
of 7\TeV at the LHC.
Events containing a single electron or muon and missing transverse momentum
are analyzed.
No significant excess of events above the standard model expectation
is found in the transverse mass distribution of the lepton-neutrino
system, and upper limits for cross sections above different transverse
mass thresholds are presented.
Mass exclusion limits at  95\% CL for a range of \Wprime models are determined, including a limit of 2.5\TeV for right-handed \Wprime bosons with standard-model-like couplings and limits of 2.43--2.63\TeV for left-handed \Wprime bosons, taking into account their interference
with the \sm $\PW$ boson.
Exclusion limits have also been set on Kaluza--Klein \WprimeKK states in the framework of split universal extra dimensions.
}

\hypersetup{%
pdfauthor={CMS Collaboration},%
pdftitle={Search for leptonic decays of W' bosons in pp collisions at sqrt(s)=7 TeV},%
pdfsubject={CMS},%
pdfkeywords={CMS, physics, W' decay}}

\maketitle %maketitle comes after all the front information has been supplied

\section{Introduction}

This Letter describes a search for a new heavy gauge boson \Wprime,
using proton-proton collision data collected during 2011 using the Compact Muon
Solenoid (CMS) detector~\cite{cms} at the Large Hadron Collider (LHC)
at a centre-of-mass energy of 7\TeV. The dataset corresponds
to an integrated luminosity of $5.0\pm0.1\invfb$~\cite{CMS-PAS-SMP-12-008}.
The search attempts to identify an excess of events with a charged lepton (an electron or muon) and a neutrino in the final state, and an interpretation of the results is provided in the context of several theoretical models.

\section{Physics models}

New heavy gauge bosons such as the \Wprime and \Zprime are predicted by various extensions of the standard model (SM).
In the sequential standard model (SSM)~\cite{reference-model}, the \Wprime boson is considered to be
a left-handed heavy analogue of the $\PW$. It is assumed to be a narrow $s$-channel resonance with decay modes and branching fractions similar to those of 
the $\PW$, with the addition of the $\cPqt\cPaqb$ channel that becomes relevant for \Wprime 
masses above 180\GeV. Interference between the \Wprime and $\PW$ is assumed to be negligible. If the \Wprime is heavy enough to decay to top and bottom 
quarks, the predicted branching fraction is about 8.5\% for each of the two leptonic channels studied in the present analysis. Under these assumptions, the 
width of a 1\TeV \Wprime is about 33\GeV.  
Decays of the \Wprime into $\PW$Z dibosons are usually suppressed in this model.

 The assumptions of the SSM were used in previous searches in leptonic channels at the Tevatron~\cite{cdf-limit,
 d0-limit} and the LHC~\cite{cms-electron-limit, cms-limit, atlas-2010, atlas-limit}. The signature of a charged
 high-momentum lepton and a neutrino would also be observed in the decays of a right-handed \WPR, predicted by
 left-right symmetric models~\cite{FourthColor,LeftRightNatur,LeftRight,Senjanovic:1978ev}.
 This particle is typically predicted to decay to a heavy right-handed
 neutrino~\cite{Min,MS,Mohapatra:1980yp}.
 
 However,
 the mass of the right-handed neutrino is not constrained, and it could be light as long as it does not couple to
 SM weak bosons. This would result in the same \WPR decay signature as for the W.

If the \Wprime is right-handed it will not interfere with the $\PW$. However, if it is left-handed (\WPL),
interference with the $\PW$ is expected
expected~\cite{criticalPaper,WprimeHelicity,dudkov}. 
Constructive (destructive) interference occurs in the mass range between $\PW$ and \Wprime
if the coupling of the \Wprime boson to quarks and leptons has opposite sign to (same sign as) the coupling of the $\PW$ boson to left-handed fermions 
($g_\mathrm{L}$). While constructive interference increases the \Wprime production cross section, 
and therefore allows experimental sensitivity at higher masses, destructive interference would yield a lower cross section, rendering previously published 
LHC mass exclusion limits ~\cite{cms-limit, atlas-limit} slightly optimistic. 
Interference has previously been considered in searches for the decay to top and bottom quarks~\cite{dudkov, tbanalysis-d0}, but never for leptonic decays.

\begin{figure}[hbtp]
\begin{center}
 \includegraphics[angle=90,width=0.58\textwidth]{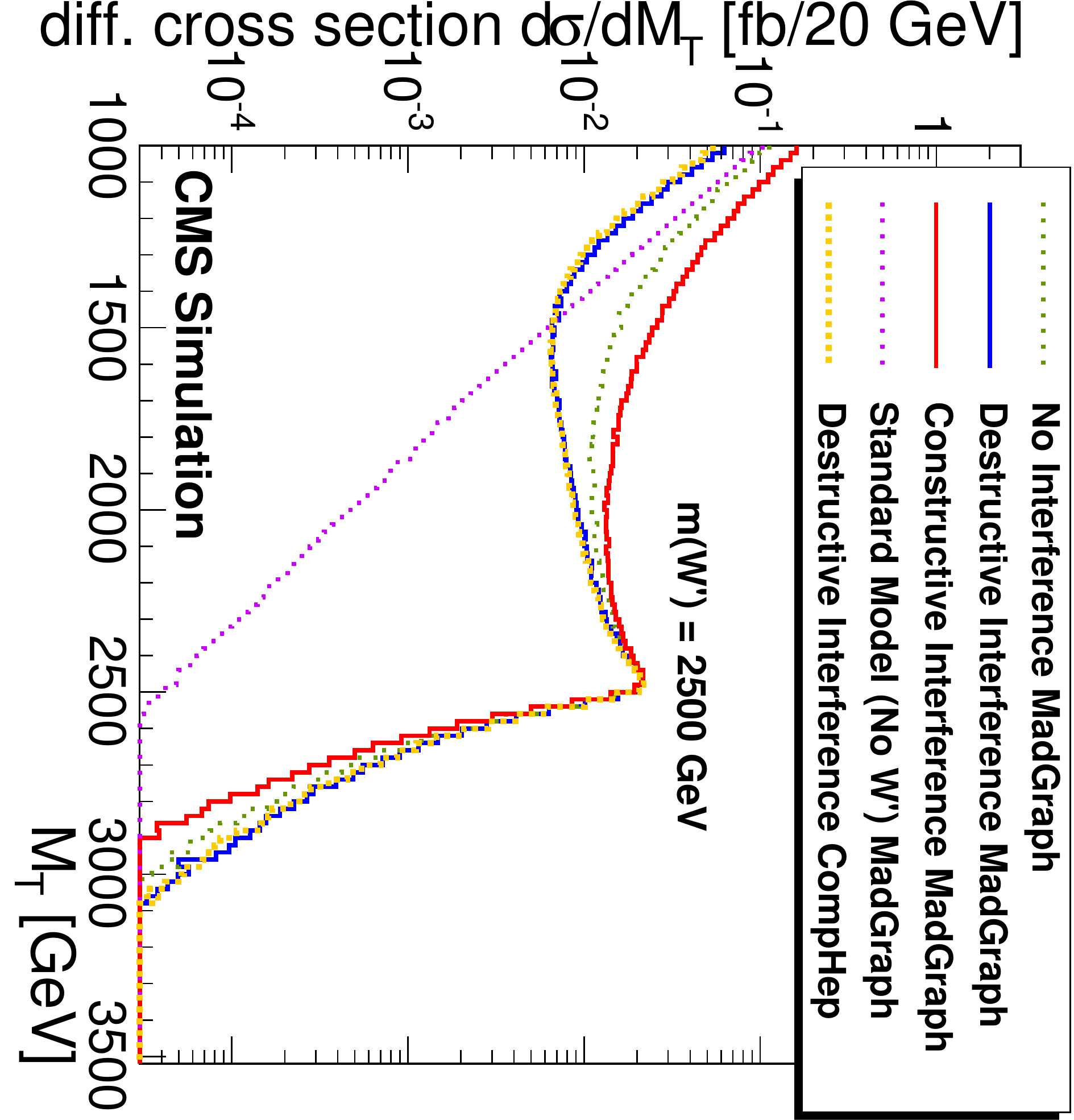}
 \caption{\MADGRAPH and \COMPHEP predictions of the transverse mass distribution for the SM $\PW$ background
and various \Wprime models for m(\Wprime)=2.5\TeV.
In the absence of interference, \WPR and \WPL cross sections are identical. A \WPL could experience constructive or
destructive interference with the SM $\PW$, yielding the shown modulation of the \MT spectrum.
}
  \label{fig:MTinterference}
\end{center}
\end{figure}

Figure~\ref{fig:MTinterference} shows the transverse mass distribution for
a \Wprime of 2.5\TeV mass
for the cases of constructive, destructive and non-interference, along with the background due to the SM $\PW$.
In the absence of interference the cross sections and transverse mass spectrum of left- and right-handed \Wprime are identical.
The \Wprime manifests itself as a Jacobian peak with its width almost independent of the presence and type of interference.
However, the intermediate region around $\MT\sim1\TeV$ shows a clear variation of the shape. Destructive interference of a \WPL boson with mass $\geq2\TeV$
modulates the W transverse mass tail, resulting in a faster fall-off.
The modulation strength and the resulting effect on the cross section both increase with the \Wprime mass and width.
Given sufficient detector resolution, the constructive and destructive interference scenarios may be distinguishable.
\\

The leptonic final states under study may also be interpreted in the
framework of universal extra dimensions (UED) with bulk mass fermions, or
split-UED~\cite{PhysRevD.79.091702, JHEP04(2010)081}.
This is a model based on an extended space-time with an additional compact
fifth dimension of radius $R$. All SM fermions and gauge bosons have Kaluza--Klein (KK) states, for instance \WprimeKKn,
where $n$ denotes the $n$-th KK excitation mode, and
\begin{eqnarray}
m_{\WprimeKKn}^2 \equiv m_n^2 = m_W^2 + \left(\frac{n}{R}\right)^2,
\label{eq:m2}
\end{eqnarray}
\begin{eqnarray}
g_n = g^{\rm SM} {\cal{F}}_n (\pi \mu R),
\label{eq:coup}
\end{eqnarray}
\begin{equation}
{\cal F}_n(x)=\begin{cases}
 0& \text{ if } n=2m+1 \\
 \frac{x^2 [-1+(-1)^m \mathrm{e}^{2x}](\coth x-1)}{\sqrt{2(1+\delta_{m0})}(x^2 + m^2\pi^2/4)}& \text{ if } n= 2m .
\end{cases}
\label{eq:fn}
\end{equation}
Here $\mu$ is the bulk mass parameter in five dimensions
of the fermion field, with $[1/R, \mu]$ defining the UED parameter space.
The coupling of the \WprimeKKn to SM fermions is denoted $g_n$ and defined as a modification of the SM coupling $g^\mathrm{SM}$ of the
$\PW$.
The function $\mathcal{F}_{2m}(x)$  tends to approach $(-1)^m \sqrt{2}$ as  $x\to \infty$.
In minimal UED models, the parameter $\mu$
is assumed to be
zero~\cite{PhysRevD.64.035002}. Following~\cite{PhysRevD.79.091702,
  JHEP04(2010)081}, we assume a non-zero value for $\mu$, thus
increasing the cross sections sufficiently to allow observation by LHC experiments.

KK-odd modes of \WprimeKKn do not couple to SM fermions, owing to KK-parity conservation. Moreover, there is no expected sensitivity for $n \geq 4$ modes at the LHC centre-of-mass energy and luminosity used in this analysis. \WprimeKKtwo is
therefore the only mode considered. Under this assumption, the decay to leptons
is kinematically identical to the sequential SM-like \Wprime
decay, and the observed limits obtained from the \WprimeENu and
\WprimeMuNu searches can directly be reinterpreted in terms of the
\WprimeKKn mass
considering the different widths.
The width of a \WprimeKKn is ${\cal F}_n^2$ times~the SSM-like \Wprime width:

\begin{equation}
\Gamma_{\mathrm{W}^\mathrm{n}_\mathrm{KK}} = {\cal F}_n^2  \frac{4}{3} \frac{m_{
\mathrm{W}^\mathrm{n}_\mathrm{KK}}}{m_{\mathrm{W}}} \Gamma_{\mathrm{W}}.
\label{eq:WKKwidth}
\end{equation}

\section{The CMS detector}

The central feature of the CMS
apparatus is a superconducting solenoid, of 6~m internal diameter, providing a magnetic field of 3.8~T.
Within the field volume are the silicon pixel and strip tracker,
the crystal electromagnetic calorimeter
and the brass/scintillator hadron calorimeter.
The electromagnetic calorimeter consists of nearly 76\,000 lead tungstate crystals.
The energy resolution for electrons with the very high transverse momentum used in this analysis, which are
predominantly in the central pseudorapidity region, is about 1\%.
In the forward region the resolution is about 2\%.
Muons are measured in gas-ionization detectors embedded in
the steel return yoke.
Central and forward regions are instrumented with four muon
stations combining high precision tracking detectors (drift tubes in the central region
and forward cathode strip chambers) with resistive plate chambers, which contribute to the trigger as well as the track measurement.
The muon transverse momentum,  $\pT^{\mu}$, is determined from the curvature of its
track, measured as it traverses the magnetized return
yoke.
Each muon track is matched to a track measured in the silicon
tracker,
resulting in a muon \pT resolution
of 1 to 10\% for \pT of up to 1\TeV.
CMS uses a two-level trigger system comprising custom hardware processors and a High-Level Trigger processor farm.
Together, these systems select around 300\unit{Hz} of the most interesting recorded bunch-crossings for permanent storage.
A detailed description of CMS can be found in Ref.~\cite{cms}.

A cylindrical coordinate system about the beam axis is used, in which the polar angle $\theta$ is
measured with respect to the counterclockwise beam direction and the
azimuthal angle $\phi$ is measured in the $xy$ plane, where the $x$ axis points towards the center of the LHC ring.
The quantity $\eta$ is the pseudo-rapidity, defined as $\eta = -\ln [\tan \theta /2]$.

\section{Event selection}
\label{sec:selection}

Candidate events with at least one high-transverse-momentum (\pT) lepton were selected using single-muon and single-electron triggers. The trigger 
thresholds were raised as the LHC luminosity increased during the data-taking period, the highest values being 
$\pT>80\GeV$ for electrons and $\pT>40\GeV$ for muons. Offline, electrons
and muons were required to have \pT at least 5\GeV higher than the online
threshold, which does not impair the search in the high mass region.

Muons were reconstructed by combining tracks from the inner tracker and the outer muon system.
Well-reconstructed muons were selected by
requiring at least one pixel hit, hits in eight tracker layers and segments in two muon stations.
Since the segments have
multiple hits and are typically found in different muon detectors
separated by thick layers of iron, the latter requirement significantly
reduces the amount of hadronic punch-through.
The transverse impact parameter $|d_0|$ of a muon track with respect to the
beam spot is required to be
less than 0.02\unit{cm}, in order to reduce the cosmic ray muon background.
Furthermore, the muon is required to be isolated within a 
$\Delta R \equiv
\sqrt{(\Delta \phi)^2 + (\Delta \eta)^2} < 0.3$ 
cone around its direction. Muon
isolation requires that
the scalar sum of the transverse momenta of all tracks originating
at the interaction vertex, excluding the muon, is less than
15\% of its \pT.
An additional requirement is that there be no second muon
in the event with \mbox{\pT $>$ 25\GeV} to
reduce the \cPZ, Drell-Yan
and cosmic ray muon backgrounds.

Electrons were reconstructed as isolated objects in the electromagnetic calorimeter,
with additional requirements on the shower shape and the ratio of hadronic to electromagnetic deposited energies.
The electrons were required to have at least one inner hit, a transverse energy greater than 85\GeV, and
required to be isolated in a cone of radius 
$\Delta R 
< 0.3$  
around the electron
candidate direction, both in the tracker and in the calorimeter.
In the tracker, the sum of the \pT of the tracks, excluding
tracks within an inner cone of 0.04, was required to be less than 5\GeV.
For the isolation using calorimeters, the total transverse energy in the barrel,
excluding deposits associated to the electron, was required to be
less than $0.03 \cdot \pt^\text{ele} + 2.0$\GeV.
The isolation requirements were
modified as luminosity
increased, owing to the increase in the typical number of additional pp
interactions (`pile-up') per LHC bunch crossing.
These
selections are designed to ensure high efficiency for electrons and a
high rejection of misreconstructed electrons from multi-jet
backgrounds.

The main observable in this search is the transverse mass \MT of the lepton-\MET system, calculated as
\begin{equation}
\MT \equiv \sqrt{2 \cdot p_\mathrm{T}^\ell  \cdot \MET \cdot (1 - \cos \Delta
  \phi_{\ell,\nu}) }
\end{equation}
where $\Delta \phi_{\ell,\nu}$ is the azimuthal opening angle between the charged lepton's transverse
momentum ($p_\mathrm{T}^\ell$) and missing transverse energy (\MET) 
direction. The neutrino is not detected directly, but gives
rise to experimentally observed \MET. This quantity was determined using a
particle-flow
technique~\cite{CMS-PAS-PFT-09-001}, an algorithm designed to reconstruct a complete list of distinct particles using all the subcomponents of the CMS detector. Muons, electrons, photons, and charged and neutral hadrons were all reconstructed individually. The \ETmiss for each event was then calculated as the vector opposing the total transverse momentum of all reconstructed particles in each event.

\begin{figure}[hbtp]
\begin{center}
 \includegraphics[width=0.49\textwidth]{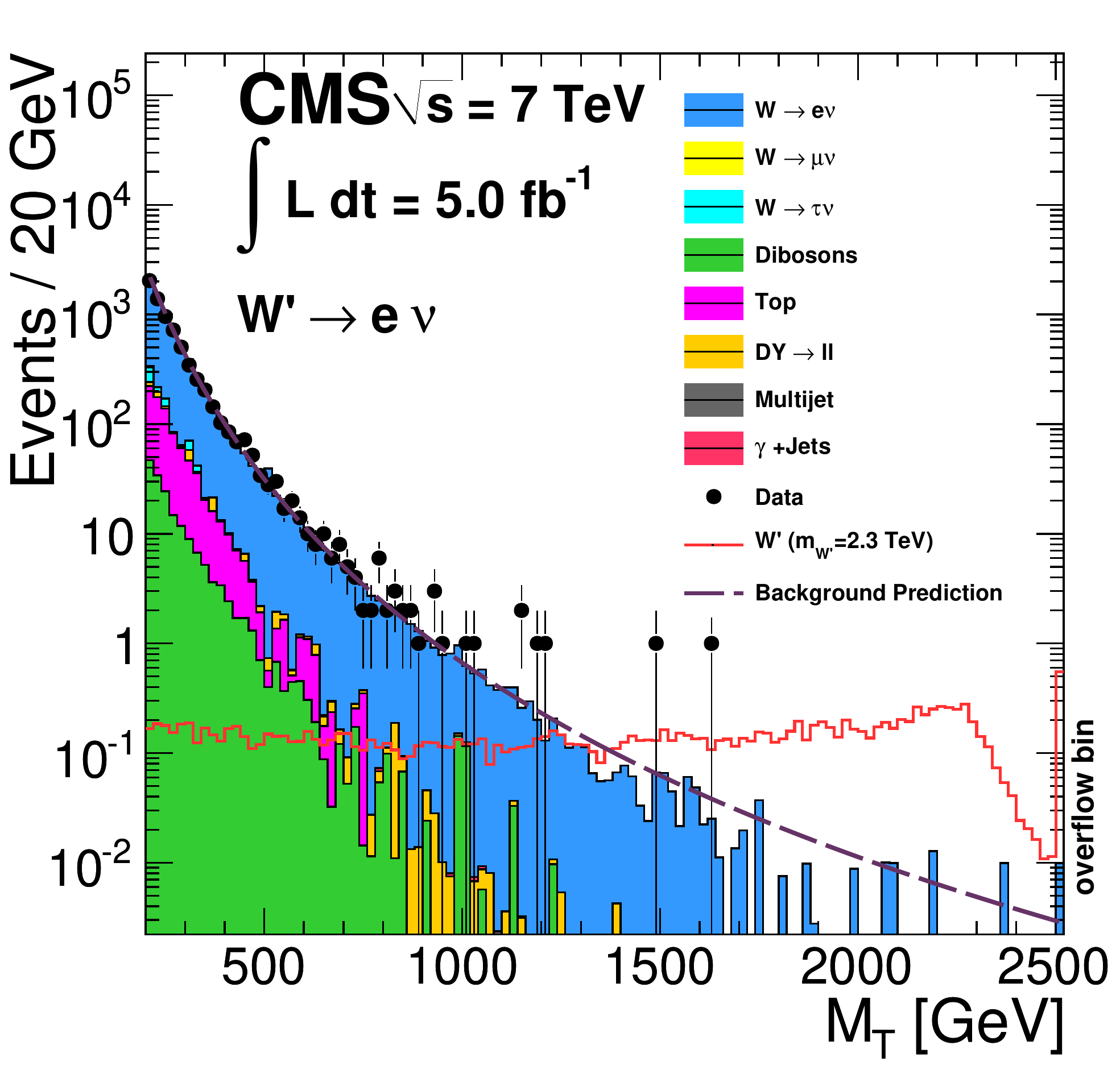}
 \includegraphics[width=0.49\textwidth]{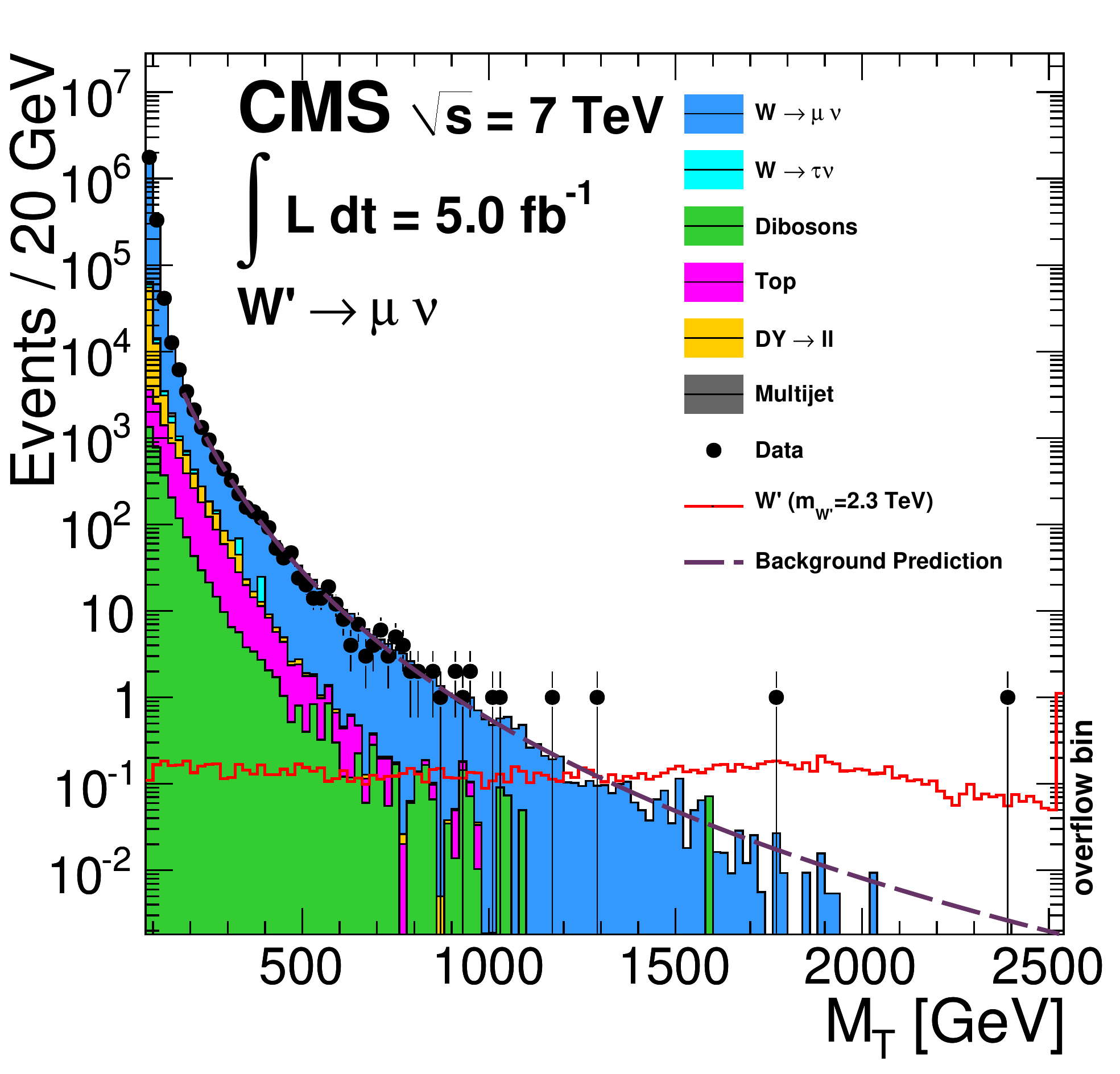}
 \caption{Observed transverse mass distributions for the electron (left) and
   muon (right) channels.
Simulated signal distributions for a (left- or right-handed) \Wprime without interference
of 2.3\TeV mass are also shown, including detector resolution
effects. The simulated background labelled as `diboson' includes WW, ZZ and WZ contributions.
The top background prediction includes single top and top pair production.
The total background prediction from a fit to the simulated transverse mass spectrum in each channel
is shown by the dashed line.
}
  \label{fig:MTboth}
\end{center}
\end{figure}

In \Wprime decays, the lepton and \ETmiss are expected to be almost back-to-back in the transverse plane,
and balanced in transverse energy. Candidate events were therefore selected through a requirement on the ratio of the lepton \pT and the \ETmiss, $0.4 <
\pT/{\ETmiss} < 1.5$. A requirement was also imposed on the angular difference in the transverse plane
of the lepton and \ETmiss direction,  $\Delta \phi_{\ell,\nu} > 0.8 \times \pi$.
No selection is made on jets.
After these selections, the average \Wprime signal efficiency for
masses up to 2.5\TeV in simulated events
was found to be around 80\% in both channels, including the roughly 90\% geometrical acceptance corresponding to a requirement
of $|\eta_{\mu}|<2.1$ for muons, and with $|\eta_{\Pe}|<1.442$ or $1.56<|\eta_{\Pe}|<2.5$ for electrons.
The transverse mass distributions after these selections are shown in Figure \ref{fig:MTboth}.

\section{Signal and background simulation}

Several large samples of simulated events were used to evaluate signal and background efficiencies.
The generated events were processed through a full simulation of the
CMS detector based on  \GEANTfour~\cite{Agostinelli:2002hh,
  Allison:2006ve}, a trigger emulation, and the event reconstruction chain.

The event samples for the \WPR signal
were produced separately from the SM W sample, using the \PYTHIA 6.4.9 generator~\cite{Sjostrand:2006za}.
This is consistent with
the case of non-interference assumed for the previous ATLAS and CMS studies.
In order to include interference of \WPL  and $\PW$ in this analysis,
a model of a single new heavy vector boson
\Wprime with a SM-like left-handed coupling strength $|g^\prime_\mathrm{L}| \approx0.65$ was implemented in the
\MADGRAPH event generator~\cite{madgraph}.
This model includes spin correlations as well as finite-width effects.
For such a left-handed scenario with interference, the generation of samples is technically more
challenging.  Since the scattering amplitude responsible for the $\ell \nu$ final state is the sum of
\WPL and
SM $\PW$ boson terms, both contributions have to be generated simultaneously.
A threshold in
\MT was applied to suppress the dominant $\PW$ contribution around the $\PW$-mass, where interference
effects
are negligible for the \WPL masses considered in this search.
The simulation uses \MADGRAPH 4.5.1, matched to \PYTHIA for showering and hadronisation.
For the hadronisation model, the \PYTHIA Tune Z2 was used for both the \WPR and \WPL simulations.
Both generators simulate at leading order (LO) and use the CTEQ6L1 parton distribution
functions (PDF)~\cite{CTEQ6L1}.
Mass-dependent K-factors, varying from 1.14 to 1.36, for the next-to-next-to-leading order (NNLO)
correction were calculated with $\textsc{fewz}$~\cite{fewz, Gavin:2012kw}. The resulting NNLO $\Wprime \to \ell
\nu$ production cross section
times branching fraction ranged from
17.7\unit{pb} (for $m_\Wprime$ = 0.5\TeV) to 0.71\unit{fb} (for $m_\Wprime= 3\TeV$) for a \Wprime without
interference (see Table~\ref{tab:ElectronMuonLimit} for cross sections).
Efficiencies and detector acceptance are then taken
into account for estimating the expected number of signal events. The
acceptance is nearly maximal since the decay 
products of such heavy particles tend to populate low
pseudorapidities. Efficiencies are high because the selections have been
optimised. 
Detailed numbers for both quantities are given in Section~\ref{sec:selection}.
The Tevatron $\WPL \to \cPqt \cPaqb$
search used the \COMPHEP generator~\cite{Comphep, ComphepManual} which has the case
of destructive interference implemented.
The agreement between the model implementations in \COMPHEP and \MADGRAPH is demonstrated for the case of destructive interference in 
Figure~\ref{fig:MTinterference}.

The primary source of background is the off-peak, high transverse mass
tail of the \sm $\PW \to \ell\nu$ decays. Other important
backgrounds arise from QCD multijet, \ttbar, and Drell--Yan events. Dibosons (WW, WZ, ZZ) decaying to
electrons, muons, or taus were also considered. The event samples for the electroweak background processes
$W \to \ell \nu$ and $\cPZ \to \ell \ell$ ($\ell = \Pe,\mu,\tau)$
were produced using \PYTHIA. NNLO cross sections were accounted for
via a single K-factor of 1.32 for the $\PW$, and mass-dependent
K-factors, ranging from 1.28 to 1.23, for the Z. The \PYTHIA generator was also used for QCD multijet events. The \ttbar
events were generated with \MADGRAPH
in combination with \PYTHIA, and the newly-calculated NNLL (next-to-leading-order including the leading logarithms of NNLO)
cross section was applied~\cite{Kidonakis:2008mu}. All other event samples were normalised to the integrated luminosity of the recorded data,
using calculated NNLO cross sections. The only exceptions were the diboson and QCD samples, for which the NLO and
LO cross sections were used respectively. We note that multijet background is largely suppressed by the event
selection requirements. The simulation of pile-up is included in all event
samples by superimposing minimum bias interactions onto the main
background processes.

In order to provide a background estimate independent of any interference effects in the W transverse
mass tail, the shape of the background was determined from simulation. The full transverse mass spectrum was
modelled by a function optimised to best describe the spectrum in either channel up to very high masses.
This function, of the form

\begin{equation}
	f(\MT)=\frac{a}{(\MT + b)^c}
\end{equation}

was fitted to the simulation and then normalised to data in the region $200\GeV < \MT<  500\GeV$,
and used to estimate the expected number of SM background events for all
transverse mass bins (shown as the dashed lines in Figure~\ref{fig:MTboth}).
A cross check under the assumption of no interference was done by fitting the \MT distribution in data confirming the simulation.
To determine the uncertainty introduced by this method, in addition to statistical errors on the fit parameters,
two alternative functions were fitted:

  \begin{equation}
    f(\MT)=\frac{a}{(\MT^2 + b \cdot \MT + c)^d}
  \end{equation}

  \begin{equation}
    f(\MT)=\frac{a(1 + \MT)^b}{(\MT^{c+d \cdot \ln \MT})}
  \end{equation}
The largest difference in the background prediction with respect to the
original fit was taken as a systematic uncertainty. For \MT larger than 1.4\TeV, this corresponds to an additional uncertainty of 0.14 events with a background expectation of 0.98 events in the muon channel and 0.26 events with a background expectation of 1.28 events in the electron channel.

\section{Systematic uncertainties}
The expected number of potential signal and background events was evaluated from simulation. In addition to uncertainties due to the fit procedure for the background, systematic
uncertainties due to imperfections in the description of the detector performance were included. Uncertainties due to the lepton energy or momentum
resolution and scale, ranging between 0.4\% and 10\%~\cite{cms-electron-limit, cms-limit} were applied to
the transverse mass spectrum.
Uncertainties due to momentum scale were evaluated using detailed
studies of the $Z\rightarrow\mu\mu$ shape and high \pT muons. The muon
\pT resolution has been previously determined with cosmic ray muons to
within 10\% for high momentum tracks~\cite{cosmics}.
In order to estimate the uncertainty on the number of expected events, the muon \pT spectrum was distorted (scaled
and smeared) according to the values extracted from comparisons with data. The missing transverse energy was
adjusted accordingly, and finally a distorted transverse mass spectrum was obtained and observed to vary by ${\sim}1\%$.
The electron energy scale uncertainty was around 1\% in the ECAL barrel and 3\% in the endcaps.
Its impact on the number of signal events above the threshold of $\MT> 600\GeV$ was ascertained to be less than
1\% for all \Wprime masses.
We assume an uncertainty of 10\%
on the hadronic component of the \ETmiss resolution (that is, excluding the lepton), and the $x$ and $y$
components of the reconstructed \ETmiss in the simulation were smeared accordingly. The impact on the number of signal events was found to be around 2\%.

Effects caused by pile-up were modeled by adding to the generated
events multiple interactions with a multiplicity distribution matched to
the luminosity profile of the collision data.
The resulting impact on the signal was studied by varying the mean of
the distribution of pile-up interactions by 8\%, yielding a variation of the signal efficiency of ${\sim}2\%$.
Following the recommendations of the PDF4LHC group~\cite{PDF4LHC}, the signal event samples for \WPR generated with \PYTHIA were reweighted using the 
LHAPDF package~\cite{LHAPDF}. PDF and $\alpha_\mathrm{s}$ variations of the
MSTW2008~\cite{Martin:2009iq}, CTEQ6.6~\cite{cteq6.6} and
NNPDF2.0~\cite{Forte:2002fg} PDF sets were taken into account and the 
impact on the signal cross sections was estimated.

\section{Results and limits}

A \WprimeENu\ or \WprimeMuNu\ signal is expected to manifest itself as an
excess over the SM expectation in the tail of the \MT distribution.
No significant excess has been observed in the data.

\begin{table}[htb]
\centering
\topcaption{\MTlower requirement for different \WPR masses,
expected number of signal and background events, number of observed events,
theoretical cross section and
upper limits on $\sigma(\WPR)\times B(\WPR
\to \ell \nu)$, with $\ell=\Pe, \mu$.
}
\label{tab:ElectronMuonLimit}
{\small
\begin{tabular}{|c|c|c|c|c|c|c|c|}
\hline
\Wprime mass & \MTlower & $N_\text{sig}$ & $N_\text{bkg}$ &
$N_\text{obs}$ & $\sigma_\text{theory}$ & Exp. Limit & Obs. Limit\\
(\GeV)         & (\GeV)         & (Events)           & (Events)           & (Events)            & (fb) & (fb)
& (fb) \\ \hline
\multicolumn{8}{|c|}{Electron channel} \\ \hline
500 & 350 & 44000 $\pm$ 4200 & 830 $\pm$ 85 & 850 & 17723 & 64.15 & 70.18 \\
700 & 550 & 9600 $\pm$ 1500 & 114 $\pm$ 15 & 128 & 4514 & 16.94 & 22.48 \\
900 & 700 & 3160 $\pm$ 460 & 37.4 $\pm$ 5.7 & 41 & 1470 & 8.38 & 9.61 \\
1000 & 800 & 1730 $\pm$ 280 & 20.0 $\pm$ 3.8 & 22 & 886 & 6.77 & 7.55 \\
1400 & 1050 & 294 $\pm$ 36 & 5.4 $\pm$ 1.6 & 6 & 144 & 3.56 & 3.77 \\
1600 & 1150 & 128 $\pm$ 13 & 3.4 $\pm$ 1.1 & 5 & 63.3 & 3.02 & 3.80 \\
1800 & 1200 & 63.9 $\pm$ 5.5 & 2.79 $\pm$ 0.99 & 3 & 28.5 & 2.53 & 2.57 \\
2100 & 1350 & 18.7 $\pm$ 1.5 & 1.55 $\pm$ 0.64 & 2 & 9.37 & 2.38 & 2.61 \\
2400 & 1450 & 5.47 $\pm$ 0.39 & 1.08 $\pm$ 0.49 & 2 & 3.40 & 2.69 & 3.39 \\
2700 & 1450 & 1.75 $\pm$ 0.13 & 1.08 $\pm$ 0.49 & 2 & 1.43 & 3.54 & 4.46 \\
3000 & 1400 & 0.59 $\pm$ 0.05 & 1.29 $\pm$ 0.56 & 2 & 0.71 & 5.45 & 6.42 \\
\hline
\multicolumn{8}{|c|}{Muon channel} \\ \hline
500 & 350 & 41000 $\pm$ 3200 & 749 $\pm$ 47 & 732 & 17723 & 44.65 & 39.13 \\
700 & 550 & 8700 $\pm$ 1000 & 102 $\pm$ 10 & 100 & 4514 & 15.42 & 14.28 \\
900 & 700 & 2920 $\pm$ 370 & 32.6 $\pm$ 5.0 & 36 & 1470 & 8.24 & 9.51 \\
1000 & 750 & 1840 $\pm$ 150 & 23.3 $\pm$ 4.2 & 26 & 886 & 6.62 & 7.57 \\
1400 & 1000 & 313 $\pm$ 25 & 5.6 $\pm$ 1.9 & 6 & 144 & 3.37 & 3.47 \\
1600 & 1100 & 136.3 $\pm$ 9.2 & 3.4 $\pm$ 1.4 & 4 & 63.3 & 2.83 & 3.04 \\
1800 & 1250 & 56.5 $\pm$ 3.7 & 1.78 $\pm$ 0.86 & 3 & 28.5 & 2.48 & 3.18 \\
2100 & 1300 & 18.5 $\pm$ 0.9  & 1.45 $\pm$ 0.75 & 2 & 9.37 & 2.35 & 2.65 \\
2400 & 1400 & 5.54 $\pm$ 0.26 & 0.98 $\pm$ 0.56 & 2 & 3.40 & 2.59 & 3.37 \\
2700 & 1450 & 1.68 $\pm$ 0.08 & 0.81 $\pm$ 0.49 & 2 & 1.43 & 3.45 & 4.77 \\
3000 & 1400 & 0.58 $\pm$ 0.03 & 0.98 $\pm$ 0.56 & 2 & 0.71 & 5.17 & 6.73 \\
\hline
\end{tabular}
}
\end{table}

For \Wprime masses well below the centre-of-mass energy of $\sqrt{s}=7\TeV$ the
signal events are expected to lie in the Jacobian peak corresponding to
the \Wprime mass. For masses above 2.3\TeV, the reduced phase space
results in many events below the Jacobian peak, and the acceptance for
the \MTlower cut drops from about 40\% for intermediate masses to 14\% at
very high \Wprime masses.
The expected signal yields given in Table~\ref{tab:ElectronMuonLimit} for a range of \WPR masses
are largely unaffected when introducing interference effects, owing to the high \MT cut corresponding to the optimum
search window, which naturally lies around the Jacobian peak.

We set upper limits on the production cross section times the
branching fraction $\sigma_{\WPR} \times \mathcal{B}(\WPR \to \ell \cPgn)$, with $\ell = \Pe$ or $\mu$.
The observed highest transverse mass events had $\MT=1.6\pm0.1\TeV$ in the electron channel,
and $\MT=2.4\pm0.1\TeV$ in the muon channel.
For $\MT>1.6\TeV$, the background expectation from the fit to simulation is less than one event in
each channel.
Cross-section limits were derived using a Bayesian method~\cite{PDGBayes} with a uniform prior probability
distribution for the signal cross section. The number of data events above an optimised transverse mass
threshold \MTlower was compared to the expected number of signal and background events.
Systematic
uncertainties on the signal and background yield were included via nuisance parameters with a log-normal prior distribution.
The \MTlower threshold was optimised for the best expected exclusion limit,
a procedure used in previous analyses~\cite{cms-limit} which is also
appropriate for establishing a \Wprime discovery.
The \MTlower threshold defining the search window increases with \Wprime mass
up to masses around 2.5\TeV, following the Jacobian peak. For larger masses, cross sections become so small
that fewer than two events are expected in the recorded data. These events are likely to have lower transverse
mass because the production is shifted to the off-peak region, as mentioned above. Both these effects serve
to lower the \MTlower threshold of the search window for very heavy \Wprime bosons.
The expected number of signal and background events listed separately for
the two channels are summarized in Table~\ref{tab:ElectronMuonLimit}. A common
theoretical NNLO cross section is assumed.

\begin{figure}[hbtp]
\begin{center}
 \includegraphics[width=0.65\textwidth]{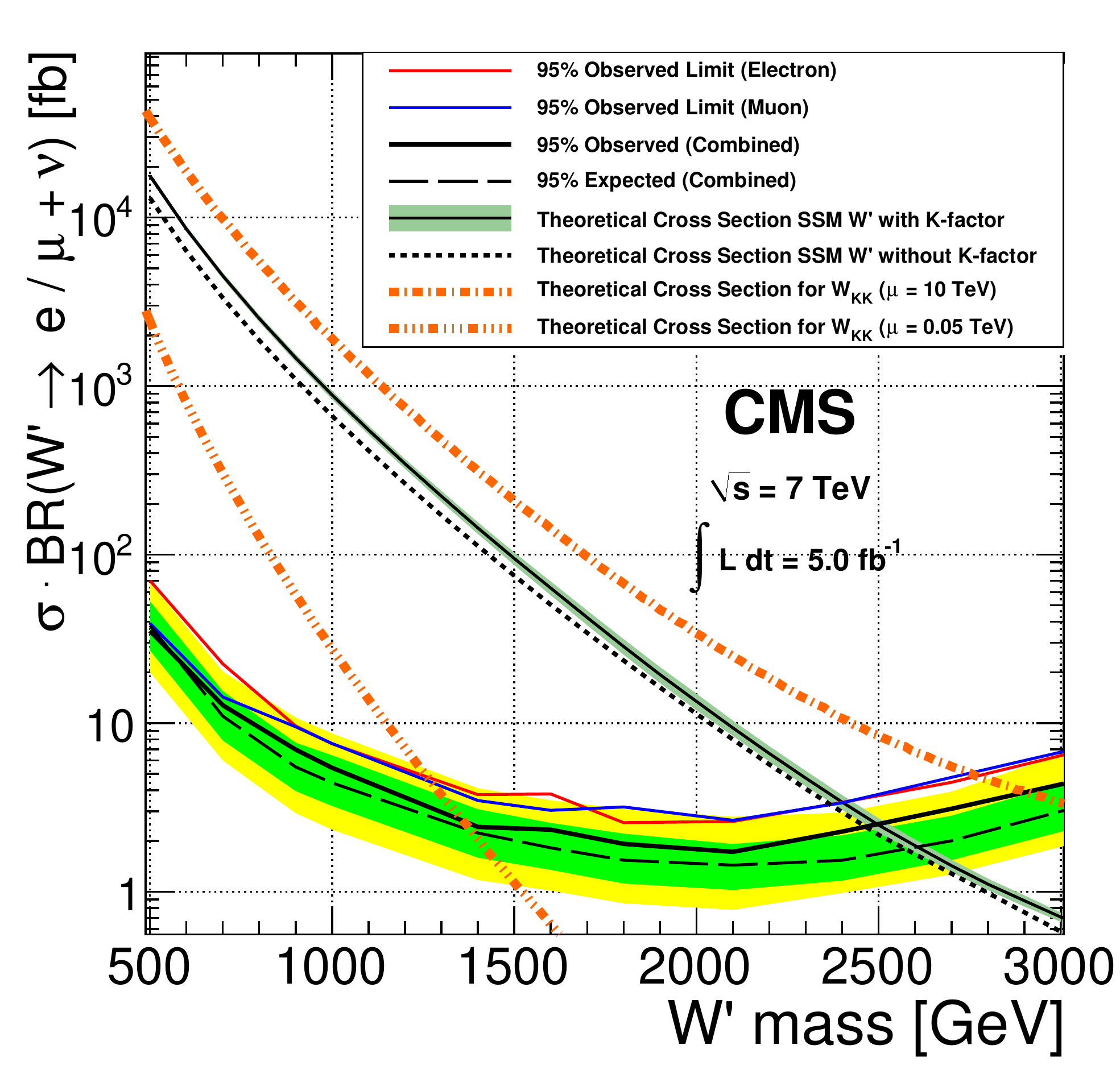}
\caption{Upper limits on
  $\sigma(\WPR)\times B(\WPR \to \ell \nu)$, with $\ell=\Pe, \mu$, and
  their combination at 95\% confidence level.
The one (two) sigma uncertainty bands are shown in green (yellow).
The theoretical cross section, with PDF uncertainties, is displayed with and
without a mass-dependent NNLO K-factor for the right-handed model without interference.
The theoretical cross sections for Kaluza--Klein \WprimeKKtwo with $\mu$=0.05\TeV and $\mu$=10\TeV are also shown.
}
\label{fig:LimitCombined}
\end{center}
\end{figure}

The expected and observed upper limits for both channels and their
combination, in the right-handed scenario without interference, are shown
in Figure~\ref{fig:LimitCombined}.
Using the central value of the theoretical cross section times the branching fraction, we exclude at 95\% confidence level (CL)
the existence of a \WPR with SM-like couplings of masses
less than 2.5\TeV (compared with an expected limit of 2.6\TeV). Note that
the background uncertainty has a negligible impact on the lower limits on
\Wprime mass, owing to the lack of observed events in the tail of the $\MT$
distribution.

A similar search procedure was performed including the effect
of interference. The theoretical cross sections
are approximately 10--30\% lower (higher) for destructive (constructive)
interference when integrating over
the transverse mass spectrum above 500\GeV  and hence influence 
the resulting mass limits~\cite{criticalPaper}.
Optimising for the best
expected cross section limit
resulted in very similar search windows at high
\MT, yielding lower limits on the \WPL mass of 2.63 (2.43)\TeV
for constructive (destructive) interference, based on the same \MADGRAPH cross sections
and K-factors as the ones used in Figure~\ref{fig:LimitCombined}. We note
that the interference affects mainly the medium \MT and hardly the
Jacobian peak region, with the latter being used to set the limits.
The limits shown do not take into account higher order electroweak
corrections at high mass, which can be sizable. The effect of these missing
corrections would be a reduction of the size of interference effects,
leading to limits that are closer to the ones quoted for the
no-interference case.

\begin{table}
\centering
\topcaption{
Excluded cross sections times branching fraction in the search window ($\MT>\MTlower$) in the electron and muon channels individually,
along with their combination. The
number of expected background events was taken from simulation. The expected and observed cross section limits are
given for each search window.}
\label{tab:LimitXsec}

\footnotesize{
\begin{tabular}{|c|c|c|c|c|c|c|c|c|c|c|}
\hline
 & \multicolumn{4}{|c|}{Electron channel} & \multicolumn{4}{|c|}{Muon channel} & \multicolumn{2}{|c|}{Combined channels} \\ \hline
$\MTlower$ & \multicolumn{2}{|c|}{Events} & \multicolumn{2}{|c|}{Limit (fb) } & \multicolumn{2}{|c|}{Events} & \multicolumn{2}{|c|}{Limit (fb) } &
\multicolumn{2}{|c|}{Limit (fb) } \\ \hline
(\GeV) & $N_\text{bkg}$ & $N_\text{obs}$  & Exp. & Obs. & $N_\text{bkg}$ & $N_\text{obs}$  & Exp. & Obs. & Exp. & Obs. \\ \hline

500 &  175 $\pm$ 22 & 192 & 10.14 & 13.85 & 158 $\pm$ 14 & 141 & 8.20 & 6.13 & 6.86 & 6.04  \\

600 &  77 $\pm$ 10 & 83 & 5.99 & 7.13 & 67.9 $\pm$ 8.1 & 62 & 5.12 & 4.46 & 4.01 & 3.95  \\

700 &  37.4 $\pm$ 5.7 & 41 & 3.80 & 4.57 & 32.6 $\pm$ 5.0 & 36 & 3.60 & 4.41 & 2.65 & 3.31  \\

800 &  20.0 $\pm$ 3.8 & 22 & 3.03 & 3.24 & 17.0 $\pm$ 3.6 & 16 & 2.95 & 2.54 & 1.94 & 1.99  \\

900 &  11.4 $\pm$ 2.6 & 12 & 2.10 & 2.30 & 9.5 $\pm$ 2.6 & 11 & 2.01 & 2.46 & 1.46 & 1.68  \\

1000 &  6.8 $\pm$ 1.8 & 8 & 1.79 & 2.02 & 5.6 $\pm$ 1.9 & 6 & 1.57 & 1.80 & 1.11 & 1.32  \\

1100 &  4.3 $\pm$ 1.3 & 6 & 1.40 & 1.88 & 3.4 $\pm$ 1.4 & 4 & 1.32 & 1.56 & 0.94 & 1.19  \\

1200 &  2.79 $\pm$ 0.98 & 3 & 1.32 & 1.32 & 2.2 $\pm$ 1.0 & 3 & 1.18 & 1.45 & 0.78 & 0.92  \\

1300 &  1.87 $\pm$ 0.74 & 2 & 1.15 & 1.15 & 1.45 $\pm$ 0.75 & 2 & 0.97 & 1.26 & 0.69 & 0.77  \\

1400 &  1.29 $\pm$ 0.56 & 2 & 0.94 & 1.22 & 0.98 $\pm$ 0.56 & 2 & 1.00 & 1.32 & 0.59 & 0.85  \\

1500 &  0.91 $\pm$ 0.43 & 1 & 0.97 & 0.97 & 0.68 $\pm$ 0.43 & 2 & 0.72 & 1.37 & 0.53 & 0.76  \\

\hline
\end{tabular}
}
\end{table}

\begin{figure}[hbtp]
\begin{center}
 \includegraphics[width=0.6\textwidth]{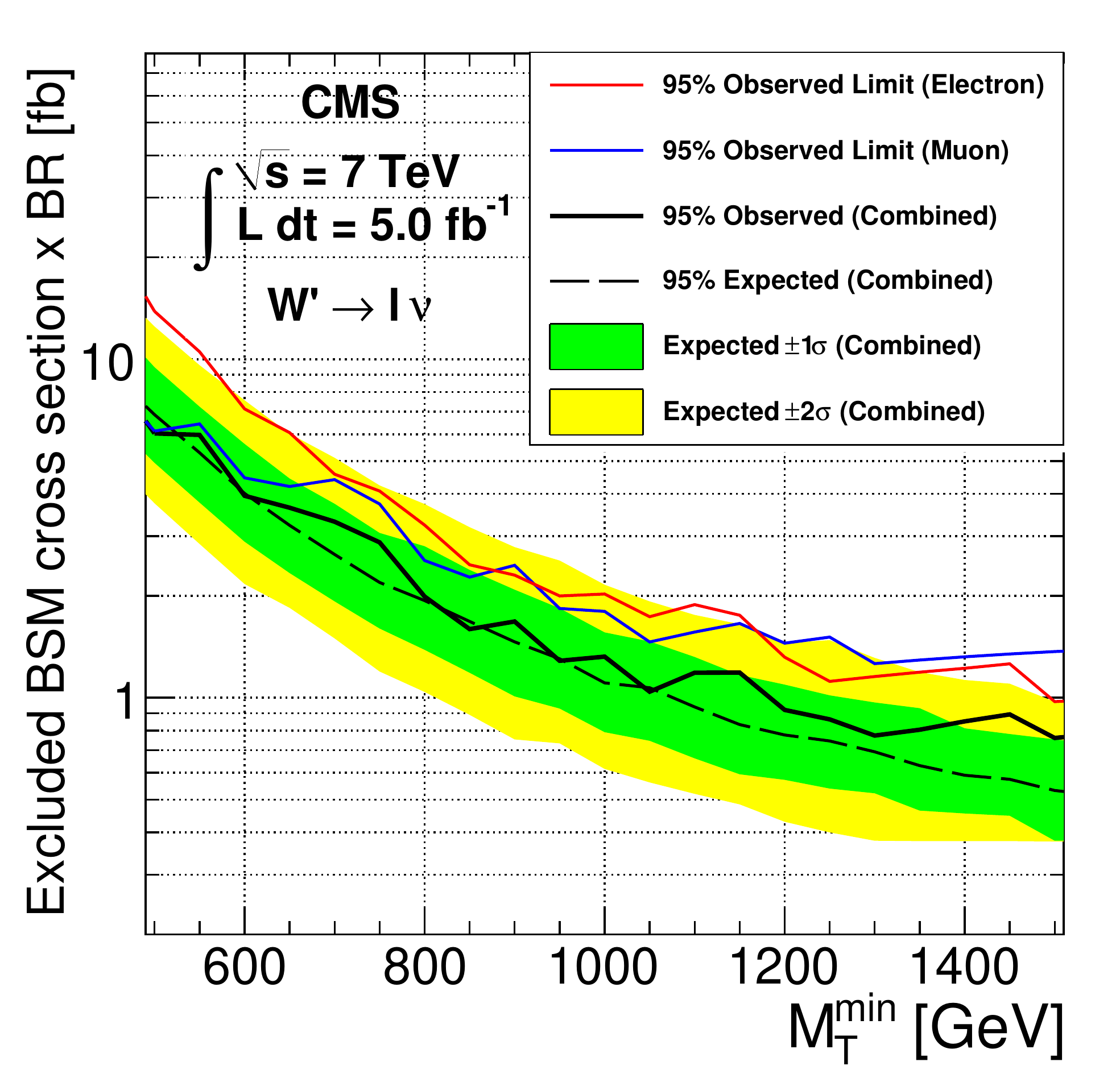}
\caption{95\% confidence level upper limits on 
the cross section times branching fraction
for physics beyond the SM (labelled BSM) 
for the charged lepton-neutrino production with transverse masses exceeding
\MTlower. The results for the electron, the muon channel, as well as for
both channels combined are presented. The one (two) sigma uncertainty
bands are shown in green (yellow).
}
\label{fig:xsec}
\end{center}
\end{figure}
In addition to the model dependent results on \Wprime production,
upper limits for the cross section of beyond-the-SM
production of charged lepton-neutrino events
are given in Table~\ref{tab:LimitXsec}
and Figure~\ref{fig:xsec}. The results are presented as a function of the
transverse mass threshold, \MTlower, and listed separately for the electron
and the muon channels, and their combination. The only assumptions made
here are that we are 
searching for a narrow $s$-channel produced resonance, using
the detector acceptance and selection efficiency outlined in
Section~\ref{sec:selection}. 
Note that the \MTlower threshold is on an
experimentally-measured quantity affected by detector resolution.

These exclusion limits on the cross-section can be translated to excluded
\Wprime masses within the context of  a given model, such as constructive or
destructive \WPL, \WPR or something else.

\begin{figure}[hbtp]
\begin{center}
 \includegraphics[width=0.6\textwidth]{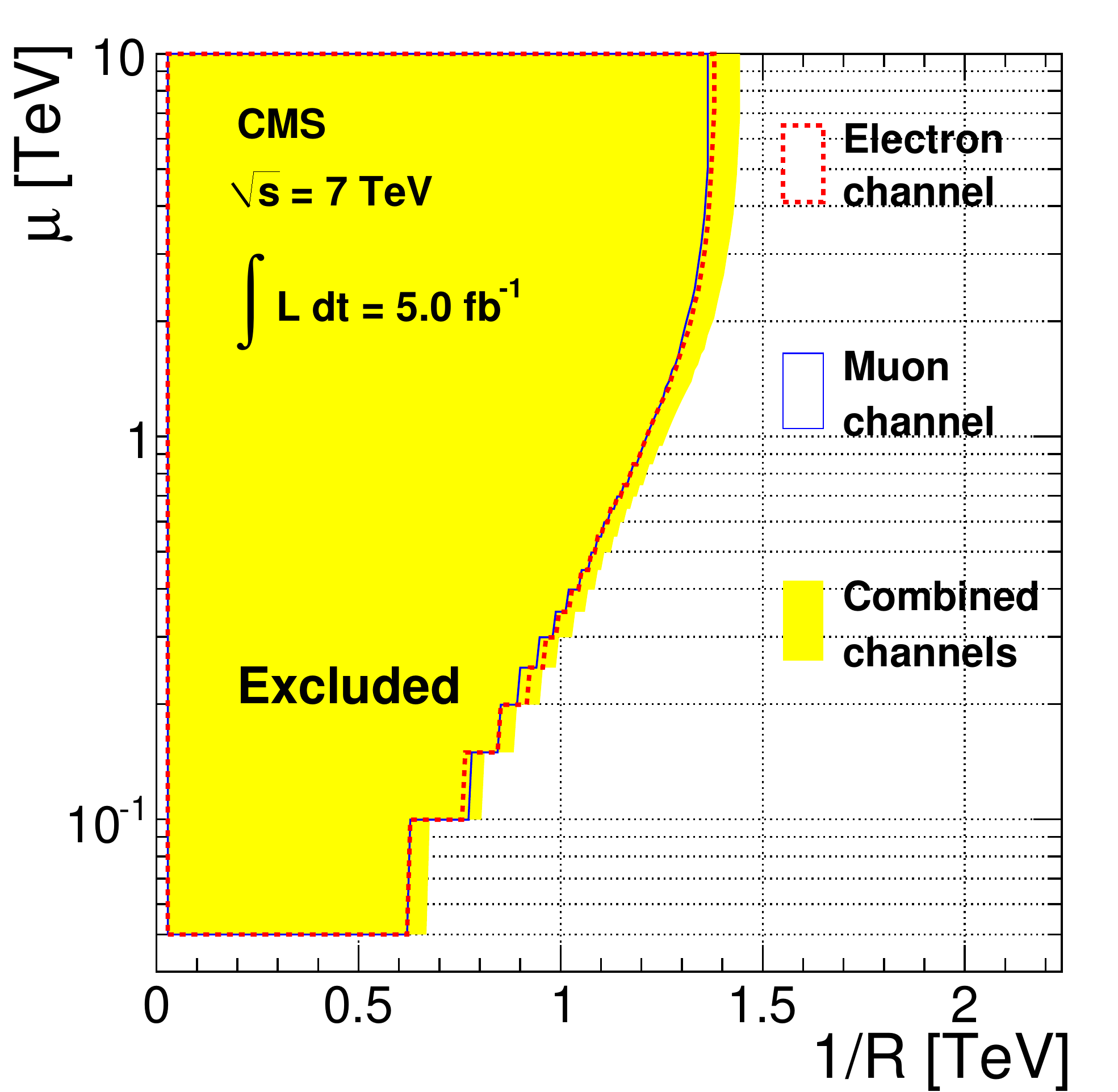}
\caption{
95\% confidence limits on the split-UED parameters $\mu$ and $R$
derived from the \Wprime mass limits taking into account the
corresponding width of the \WprimeKKtwo. The colored areas correspond to the
\WprimeKKtwo exclusion regions with the same final state as the SM-like
\Wprime. Results are shown for the electron and muon channels, as well as
for both channels combined. The \WprimeKKtwo is the lowest state that can
couple to SM fermions. Since it has even parity it can be produced
singly.
}
\label{fig:UED}
\end{center}
\end{figure}

The observed limits illustrated in Figure~\ref{fig:LimitCombined} can be reinterpreted in terms of the
\WprimeKKtwo mass, as shown in the same figure for values of the bulk mass
parameters $\mu$ = 0.05\TeV
and $\mu$ = 10\TeV.
For these parameters the second Kaluza--Klein excitation \WprimeKKtwo has been excluded for masses below 
1.4\TeV ($\mu$ = 0.05\TeV) or 2.9\TeV ($\mu$ = 10\TeV), respectively.
The corresponding widths (Eq.~\eqref{eq:WKKwidth})
are taken into account in the calculation of the cross section
times the branching fraction of \WprimeKKtwo.
These lower limits on the mass can be directly translated to bounds on the
split-UED parameter space $[1/R, \mu]$ with
$\mu$ being the mass parameter for bulk fermions and $R$ the radius of the extra dimension.
The results are displayed in Figure~\ref{fig:UED}, using the relations between $R$, $\mu$ and the \WprimeKKtwo
mass, and the couplings to SM fermions described by expressions \eqref{eq:m2}, \eqref{eq:coup} and \eqref{eq:fn}.
The split-UED model also allows for $\PW$-\Wprime interference.
When the constructive case is considered, it has a comparable
sensitivity to the no-interference case.

\section{Summary}

A search for an excess of events with a final state consisting of a charged lepton (electron or muon) and significant 
missing transverse momentum has been performed, using 5.0\invfb of $\sqrt{s}$ = 7\TeV pp collision data. No significant 
excess over the SM expectation was observed in the distribution of transverse mass. A \WPR in the SSM with a mass of less 
than 2.5\TeV has been excluded at 95\% CL. For the first time in such a study, $\PW$-\Wprime interference effects have 
been taken into account, and mass exclusion limits have been determined as 2.63\TeV and 2.43\TeV for constructive and 
destructive interference respectively. These are the most stringent limits yet published.      
An interpretation of the search results has also been made in a specific framework of universal extra dimensions with 
bulk mass fermions. 
The second Kaluza--Klein excitation \WprimeKKtwo has been excluded for masses below 
1.4\TeV, assuming a bulk mass parameter $\mu$ of 0.05\TeV or masses below 2.9\TeV for $\mu$=10\TeV.

\clearpage

\section*{Acknowledgements}

We congratulate our colleagues in the CERN accelerator departments for the excellent performance of the LHC 
machine. We thank the technical and 
administrative staff at CERN and other CMS institutes, and acknowledge support from: FMSR (Austria); FNRS 
and FWO (Belgium); CNPq, CAPES, FAPERJ, and
FAPESP (Brazil); MES (Bulgaria); CERN; CAS, MoST, and NSFC (China); COLCIENCIAS (Colombia); MSES (Croatia); 
RPF (Cyprus); MoER, SF0690030s09 and ERDF 
(Estonia); Academy of Finland, MEC, and HIP (Finland); CEA and CNRS/IN2P3 (France); BMBF, DFG, and HGF 
(Germany); GSRT (Greece); OTKA and NKTH (Hungary); 
DAE and DST (India); I$\pm$ (Iran); SFI (Ireland); INFN (Italy); NRF and WCU (Korea); LAS (Lithuania); 
CINVESTAV, CONACYT, SEP, and UASLP-FAI (Mexico); MSI 
(New Zealand); PAEC (Pakistan); MSHE and NSC (Poland); FCT (Portugal); JINR (Armenia, Belarus, Georgia, 
Ukraine, Uzbekistan); MON, RosAtom, RAS and RFBR 
(Russia); MSTD (Serbia); MICINN and CPAN (Spain); Swiss Funding Agencies (Switzerland); NSC (Taipei); TUBITAK 
and TAEK (Turkey); STFC (United Kingdom); DOE 
and NSF (USA).  
Individuals have received support from the Marie-Curie programme and the European Research Council (European 
Union); the Leventis Foundation; the A. P. 
Sloan Foundation; the Alexander von Humboldt Foundation; the Belgian Federal Science Policy Office; the Fonds 
pour la Formation \`a la Recherche dans 
l'Industrie et dans l'Agriculture (FRIA-Belgium); the Agentschap voor Innovatie door Wetenschap en 
Technologie (IWT-Belgium); the Council of Science and 
Industrial Research, India; and the HOMING PLUS programme of Foundation for Polish Science, cofinanced from 
European Union, Regional Development Fund.

\bibliography{auto_generated}   % will be created by the tdr script.

\cleardoublepage \appendix\section{The CMS Collaboration \label{app:collab}}\begin{sloppypar}\hyphenpenalty=5000\widowpenalty=500\clubpenalty=5000\input{EXO-11-024-authorlist.tex}\end{sloppypar}
\end{document}

%% file: EXO-11-024-authorlist.tex
\textbf{Yerevan Physics Institute,  Yerevan,  Armenia}\\*[0pt]
S.~Chatrchyan, V.~Khachatryan, A.M.~Sirunyan, A.~Tumasyan
\vskip\cmsinstskip
\textbf{Institut f\"{u}r Hochenergiephysik der OeAW,  Wien,  Austria}\\*[0pt]
W.~Adam, T.~Bergauer, M.~Dragicevic, J.~Er\"{o}, C.~Fabjan, M.~Friedl, R.~Fr\"{u}hwirth, V.M.~Ghete, J.~Hammer\cmsAuthorMark{1}, N.~H\"{o}rmann, J.~Hrubec, M.~Jeitler, W.~Kiesenhofer, V.~Kn\"{u}nz, M.~Krammer, D.~Liko, I.~Mikulec, M.~Pernicka$^{\textrm{\dag}}$, B.~Rahbaran, C.~Rohringer, H.~Rohringer, R.~Sch\"{o}fbeck, J.~Strauss, A.~Taurok, F.~Teischinger, P.~Wagner, W.~Waltenberger, G.~Walzel, E.~Widl, C.-E.~Wulz
% ...
\vskip\cmsinstskip
\dag:~Deceased\\
1:~~Also at CERN, European Organization for Nuclear Research, Geneva, Switzerland\\
% ...
